\documentclass[journal]{IEEEtran}
\usepackage{cite}
\usepackage{amsmath,amssymb,amsfonts}
\usepackage{algorithmic}
\usepackage{graphicx}
\usepackage{textcomp}
\usepackage{leftidx}
\usepackage{multirow}
\usepackage[dvipsnames]{xcolor}
\usepackage[ruled,vlined,linesnumbered]{algorithm2e}
\renewcommand{\baselinestretch}{0.9575}
\ifCLASSINFOpdf
\else
\fi

\begin{document}

\title{H$_2$O-Cloud: A Resource and Quality of Service-Aware Task Scheduling Framework for Warehouse-Scale Data Centers - A Hierarchical Hybrid DRL (Deep Reinforcement Learning) based Approach}

\author{Mingxi Cheng, Ji Li, Paul Bogdan,~\IEEEmembership{Member,~IEEE}, and
        Shahin Nazarian
\thanks{The authors are with the Department of Electrical and Computer Engineering, University of Southern California, Los Angeles, CA, 90089, USA
(e-mail: mingxic@usc.edu; jli724@usc.edu; pbogdan@usc.edu; shahin.nazarian@usc.edu).}}

\markboth{Journal of \LaTeX\ Class Files,~Vol.~14, No.~8, August~2015}%
{Shell \MakeLowercase{\textit{et al.}}: Bare Demo of IEEEtran.cls for IEEE Journals}

\maketitle

\begin{abstract}
Cloud computing has attracted both end-users and Cloud Service Providers (CSPs) in recent years.
Improving resource utilization rate (RUtR), such as CPU and memory usages on servers, while maintaining Quality-of-Service (QoS) is one key challenge faced by CSPs with warehouse-scale datacenters.
Prior works proposed various algorithms to reduce energy cost or to improve RUtR, which either lack the fine-grained task scheduling capabilities, or fail to take a comprehensive system model into consideration.
This article presents H$_2$O-Cloud, a Hierarchical and Hybrid Online task scheduling framework for warehouse-scale CSPs, to improve resource usage effectiveness while maintaining QoS.
H$_2$O-Cloud is highly scalable and considers comprehensive information such as various workload scenarios, cloud platform configurations, user request information and dynamic pricing model.
The hierarchy and hybridity of the framework, combined with its deep reinforcement learning (DRL) engines, enable H$_2$O-Cloud to efficiently start on-the-go scheduling and learning in an unpredictable environment without pre-training.
Our experiments confirm the high efficiency of the proposed H$_2$O-Cloud when compared to baseline approaches, in terms of energy and cost while maintaining QoS. 
Compared with a state-of-the-art DRL-based algorithm, H$_2$O-Cloud achieves up to $201.17\%$ energy cost efficiency improvement, $47.88\%$ energy efficiency improvement and $551.76\%$ reward rate improvement.
\end{abstract}


\IEEEpeerreviewmaketitle

\section{Introduction} \label{intro}

\IEEEPARstart{A} cloud is a platform over network where information technology and computing resources are available instantly and on-demand \cite{buyya2010special}.
Cloud computing has become an efficient and powerful archetype for both end-users and cloud operators. 
The advantages of high computing power, low service cost, high scalability and availability attract end-users with desirability to implement computational and power consuming applications at low expense.
The huge revenue opportunity is an inviting treasure for Cloud Service Providers (CSPs).
Recent reports have shown that the global public cloud market revenue was more than \$$130$ billion in 2017 and is expected to exceed \$$500$ billion in 2026.
Google App Engine (GAE) \cite{googleGAE}, 
Amazon Web Service (AWS) \cite{amazonaws}, and Windows Azure \cite{Windowsazure} are among the popular commercial services in cloud computing.

Although the revenue of cloud computing is notable, the electricity bill of data centers (DCs) is also noteworthy. DCs in US alone used $91$ billion kilowatt-hours of electrical energy in 2013, and the annual usage will increase to $139$ billion kilowatt-hours by 2020, with the cost of power of \$$13.7$ billion \cite{dcpolluters}.
With large-scale DCs, resource management of cloud-platform enables CSPs to achieve economies of scale by improving cost effectiveness \cite{delimitrou2014quasar}.
However, power efficiency of large-scale DCs is harmed by inefficient resource utilization of servers, which is a well-known problem faced by modern CSPs \cite{reiss2012heterogeneity,barroso2011warehouse,reiss2012towards}. 

CPU and memory utilization of a Google cluster with $12.5$k machines in one month has been analyzed by former researchers \cite{reiss2012towards}. 
The majority utilization rate of both CPU and memory resources is below $50\%$ for most of the time, which harms power usage effectiveness \cite{kozyrakis2013resource}. 
Similarly, Barroso et al. \cite{barroso2013datacenter} note
two critical features of energy usage of DCs: (i) servers tend to be more energy inefficient under low utilization rate (with the optimal power efficient utilization rate of most servers ranging between $70\%$ and $80\%$), and (ii) servers consume a considerable amount of power in idle mode. 
Therefore, resource-efficient cluster management should take DCs' resource utilization rate (RUtR) into consideration while maintaining Quality of Service (QoS).
QoS is defined based on a combination of task rejection rate, deadline violation rate and reward rate (which measures how often the user-designated task priority requirement is fulfilled) in this work.
For profit-driven CSPs in industrial community with warehouse-scale DCs, effective cloud resource management and task scheduling are resultful measures to broaden profit margin.

\begin{figure}[!t]
	\centerline
	{\includegraphics[width=\columnwidth]{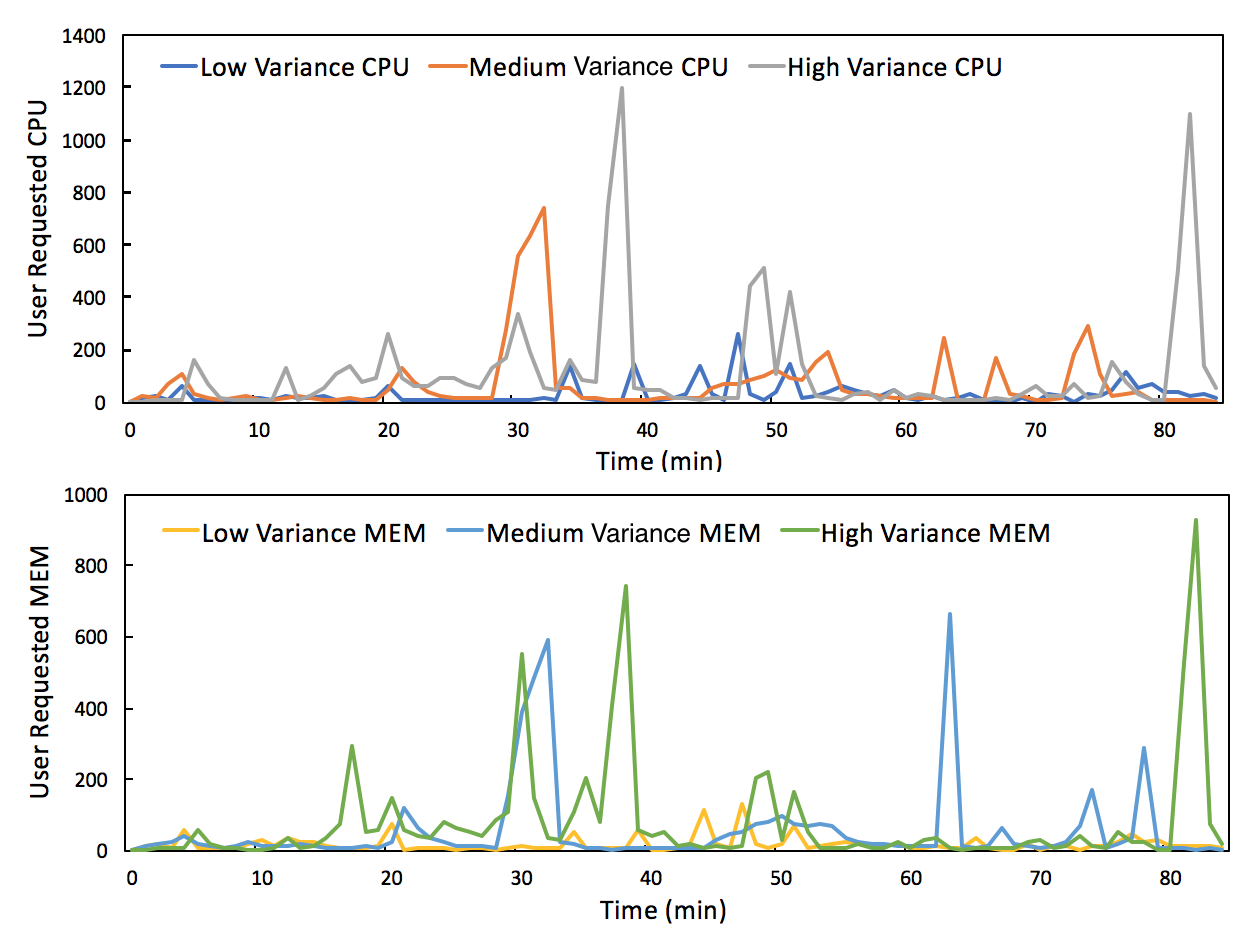}}
	\vskip -4mm
	\caption{Analysis of CPU and memory resource requirements for a Google cluster with $12.5$k machines over $1.5$ hour. Three types of user workload patterns are shown: low variance, medium variance, high variance in both CPU and memory (MEM).
    }
	\label{fig:user_request}
	\vskip -5mm
\end{figure}

Along with the fast-growing cloud computing market, a number of challenges have arisen with respect to resource management and task scheduling: 
(i) Scalability has become an important consideration to support large-scale DCs that have been investigated or are being planned worldwide.
(ii) Self-learning techniques are required to handle the uncertainty and changing nature of user requests \cite{delimitrou2016hcloud, yao2019tvlsi}.
Figure \ref{fig:user_request} shows that user requests have different patterns, and resource requirements such as CPU and memory vary over time. 
(iii) Short runtime is highly desirable, in order to provide quick response to user requests.
To deal with these challenges, intelligent cloud resource management has risen and become the state-of-the-art fashion in cloud computing \cite{zhang2018intelligent}.

In this work, we propose a \underline{H}ierarchical and \underline{H}ybrid \underline{O}nline task scheduler for \underline{Cloud} computing (H$_2$O-Cloud), inspired by deep reinforcement learning (DRL) \cite{mnih2015human}.
DRL is effective for problems where state space has high dimensions and action space is rather limited \cite{liu2017hierarchical,mnih2015human}. Therefore, the large dimension of state space (i.e., large-scale cloud platform configuration) in this cloud resource management and task scheduling problem can be resolved by applying DRL technique.
In order to handle the enormous action space (i.e., a large number of available servers and exact task execution time options) and realize real-time scheduling, a hierarchical task scheduling process is proposed.
More precisely, the proposed hierarchical method decouples the action space into several sub-action spaces: server farms, servers, hours and minutes, where the decision for each sub-action is made by one Deep Q-Network (DQN).
The specialized DQNs not only schedule tasks faster than one combined large DQN, but also improve the decision quality as one DQN is responsible for one type of action.

The proposed hierarchical task scheduling process schedules an incoming task into a server farm, and chooses a server in selected server farm; then takes dynamic pricing model (proposed in Section \ref{pricemodel}) into consideration to select optimal hour and minute(s) to run the task (details presented in Section \ref{thesystem} and Section \ref{training}). 
In the proposed framework, uncertainties in user requests are handled by the DQNs that are capable of autonomic learning. In addition, the fast decision-making challenge as well as the scalability are enabled by the online scheduling of DRL. 
Comprehensive environment information is considered in our model, such as user-request information, current environment status, historical decisions, user workload patterns and realistic electric price model, etc. Our proposed method can be extended to include more input information as either new features of certain DQNs or new DQN layers. 

In summary, the main contributions of our paper are as follows.
\begin{itemize}
\item \textbf{Four-Layer Hierarchical Architecture.}
H$_2$O-Cloud schedules incoming user requests in series and makes complex long-term decisions hierarchically. The four-layer structure is trained individually on-the-go without pre-training. 
It breaks down massive state and action spaces generated by the warehouse-scale cloud platform configuration, hence speeds up the decision-making process while maintaining high performance.
\item \textbf{Hybrid Mechanism.} 
The hybrid mechanism provides valid actions when DRL agent generates invalid actions in some corner cases. This back-up mechanism further reduces the task rejection rate and contributes to improvement of QoS, which is defined earlier as the task rejection rate, deadline violation rate and reward rate. 
\item \textbf{Fine-grained Online Scheduling for Warehouse-Scale CSPs.}
This work achieves fine-grained (minute-level scheduling) high-performance online scheduling for warehouse-scale CSPs by using DRL without the need for pre-training.
\end{itemize}

The remainder of this paper is organized as follows: related work is reviewed in Section \ref{relatedwork}, Major properties of H$_2$O-Cloud and system model are presented in Section \ref{systemmodel}, The structure and algorithm of H$_2$O-Cloud are presented in Section \ref{overviewofsystem} and Section \ref{drlandcontrol} respectively, and the experiment results are discussed in Section \ref{experiment}. 

\begin{figure*}
	\centering
	\includegraphics[width=\textwidth]{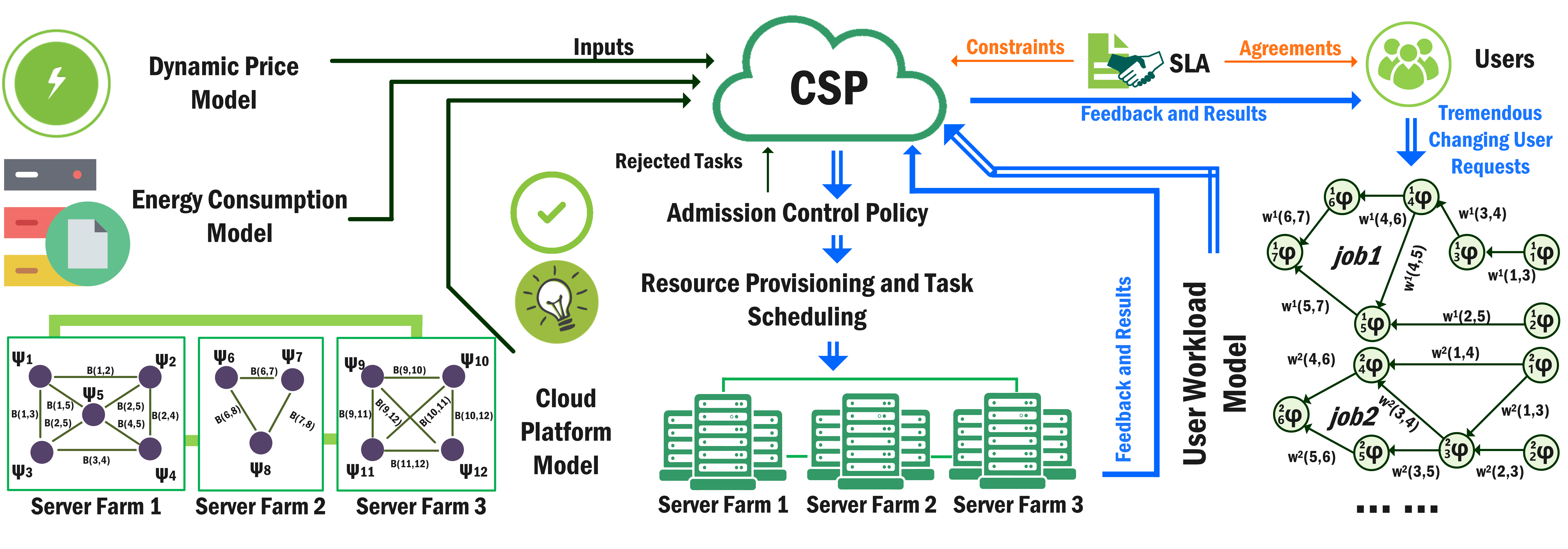}
	\vskip -5mm
	\caption{H$_2$O-Cloud system models of cloud platform and structure, including cloud platform model, user workload model, energy consumption model, and dynamic price model defined in Sections \ref{cloud_platform_model} to \ref{pricemodel}.}
    \label{fig:drlsystem}
    \vskip -5mm
\end{figure*}

\section{Related Work} \label{relatedwork}

Cloud resource management has become a popular research area and many approaches have been proposed to fulfill various objectives. The authors in \cite{kozyrakis2013resource,delimitrou2013paragon} proposed Paragon to increase the capability and cost effectiveness of large-scale DCs by improving their resource efficiency. 
Delimitrou et al. proposed a resource-efficient and QoS-aware cluster management system for joint resource allocation and assignment  \cite{delimitrou2014quasar}. 
The authors in \cite{delimitrou2015tarcil} presented a distributed scheduler that targets both scheduling speed and quality, also performs admission control. 
A hybrid cloud resource provisioning system that uses both reserved and on-demand resources was proposed in \cite{delimitrou2016hcloud}. 
Nevertheless, these prior works scale up to $1000$ servers which is not enough for a warehouse-scale CSP (with $10,000+$ servers \cite{barroso2013datacenter}).
The authors in \cite{zhang2013harmony,gao2013energy,xiaoguang2013research,li2017fast} utilized a realistic electric price model to improve the energy cost efficiency; however their solution suffers from scalability issues, and their offline algorithms have difficulties in dealing with the large size of inputs.
N. Liu et al. applied DRL techniques to (partially) solve the resource allocation problem \cite{liu2017hierarchical}, which can deal with scalability issues. Their solution, however, does not schedule tasks with data dependencies, which is critical to guarantee that parallel tasks are executed correctly in the cloud environment \cite{isard2007dryad}. 
The authors in \cite{wei2018drl} proposed a DRL-based QoS-aware job scheduling framework to reduce average job response time, where QoS requirement is defined as the expected job response time. However, energy cost is not considered in their work.
None of these works conducts fine-grained online scheduling, i.e., scheduling each task into specific minute(s) on-the-fly while targeting energy cost reduction at the same time.

\section{H$_2$O-Cloud: Properties and System Modeling} \label{systemmodel}
The properties of H$_2$O-Cloud and system model are introduced in this section, including the models for user workload, cloud platform, energy consumption and electricity price.

\subsection{Property} \label{property}
\subsubsection{Resource-Efficient}
CSPs that own warehouse-scale DCs are incentivized by increasing the profit through minimizing energy cost while meeting users' requirements in cloud services. 
Utilizing resources efficiently is a reasonable way of achieving economies of scale.
Considering a server tends to be more energy-efficient when its RUtR lies in the optimal range, but consumes a lot of energy when idle \cite{barroso2013datacenter}, the resource efficiency of H$_2$O-Cloud is realized through scheduling tasks intelligently to make each running server work at its optimal energy-efficient operating range and to selectively shut down idle servers.

\subsubsection{QoS-Aware}
Task priority and deadline describe the importance and emergency degree of a task, which are designated by end-users.
In the case of contention, tasks with higher priorities should be executed before tasks with lower priorities, and tasks with earlier deadlines should be executed before tasks with loose deadlines.
Awareness of task priorities and deadlines enables H$_2$O-Cloud to achieve priority-aware task scheduling with low deadline violation rate and to yield better QoS.
In addition, the commonly used soft-deadline for tasks \cite{kondo2009cost, mao2011auto, mattess2013scaling} is adopted in this work. 
In contrast to a hard-deadline, the missing of which causes permanent task rejection and is defined as catastrophic failure, a soft-deadline allows a task to be recycled and rescheduled in case of a deadline violation. 
Utilization of soft-deadlines equips the presented framework with a solution to reduce task rejection rate while guaranteeing admission control. 

\subsubsection{Fast, Warehouse-Scale, Online}
H$_2$O-Cloud allocates resources and schedules tasks by using the DRL-based hybrid algorithm, which enables quick effective resource allocation and task scheduling even on warehouse-scale cloud platforms.
H$_2$O-Cloud is able to make decisions on-the-fly, given that it is aware of historical decisions and current utilization of all DCs, and does not need to wait for the presence of all task requests to make high quality decisions.
Fast convergence of DRL algorithm gives good performance in runtime, cost efficiency, and energy efficiency without pre-training the models.

\subsection{Cloud Platform Model} \label{cloud_platform_model}
Virtualization is the fundamental technology of cloud computing, which enables multiple operating systems to run on the same physical platform \cite{rittinghouse2016cloud}.
One server in cloud platform can hold several virtual machine (VM) configurations, and servers can hold different machine types.
VMs are used by CSPs to provide infrastructures, platforms, and resources (e.g., CPU, memory, storage, etc.) to run end-users' applications.
For example, Google App Engine (GAE) supports more than twenty machine types to configure cloud services differently, such as high-memory configuration and high-CPU configuration \cite{googlemachinetype}.

Graph modeling is used to present server farms in this work.
Each vertex in the graph represents a server, and each edge represents a channel between servers.
As shown in Figure \ref{fig:drlsystem}, a CSP owns $N$ servers $\{\psi_1, \psi_2,..., \psi_{N}\}$, and nearby servers are structured into $M$ clusters (or farms, server cluster and server farm are equivalent in this work). 
Server clusters are connected with two-way high-speed channels, whereas servers within one server cluster are connected through local channels.
The bandwidth of channels between servers $\psi_i$ and $\psi_j$ in a cluster is represented by the weight of the corresponding edge (denoted by $B(i,j)$).
A CSP supports $V$ types of VMs $\{VM_1, ...,VM_V\}$ in total, and each server supports a subset of these types.
Each type $VM_v$ ($v\in[1,V]$) is associated with a two-tuple parameter set $\{R^v_{CPU},R^v_{MEM}\}$, which represents the required amount of CPU and memory units, respectively. 

A server $\psi_n$ holds in total $CPU_n$ units of CPU and $MEM_n$ units of memory. Therefore, at any time $t$, the total amount of CPU and memory structured in VMs on server $\psi_n$  should be less than the upper bound of its CPU and memory, i.e., 
\begin{equation} 
  \Sigma_{v\in V_n} R_{CPU}^v \leq CPU_n,
\end{equation}
\begin{equation} 
  \Sigma_{v\in V_n} R_{MEM}^v \leq MEM_n.
\end{equation}

\subsection{User Workload Model} \label{workloadmodel} 
In this paper, the entire user workload is composed of a number of \textit{jobs} (i.e., user requests), each of which containing several \textit{tasks} with dependencies.
A job is modeled by a
Directed Acyclic Graph (DAG) \cite{warneke2009nephele,isard2007dryad}, as shown in Figure \ref{fig:drlsystem}.
A vertex ${^u}{_i}\varphi$ in the DAG represents the $i$-th task in the $u$-th job. 
The weight of an edge $w^u(a, b)$ indicates the amount of dependent data that needs to be transferred from task ${^u}{_a}\varphi$ to task ${^u}{_b}\varphi$ in $job_u$. 
Using $job_1$ in Figure \ref{fig:drlsystem} as an example: $job_1$ contains $7$ tasks $\{{^1}{_1}\varphi,...,{^1}{_7}\varphi\}$, and edge weight $w^1(a,b)$ ($a,b \in [1,7]$) is the amount of data that needs to be transferred from task ${^1}{_a}\varphi$ to task ${^1}{_b}\varphi$ so that the following task ${^1}{_b}\varphi$ can be executed.

Each task ${^u}{_i}\varphi$ (i.e., task $i$ in $job_u$) is associated with a parameter tuple $\{{^u}{_i}D_{CPU},{^u}{_i}D_{MEM},{^u}{_i}D_{VM},{^u}{_i}Prr,{^u}{_i}DDL\}$, which represents the requested CPU, memory, VM type, priority and soft-deadline of task ${^u}{_i}\varphi$, respectively. 
Each VM type has its own amount of CPU and memory, i.e.,
\begin{equation} 
	{^u}{_i}D_{CPU} = [{^u}{_i}D_{CPU}^1,...,{^u}{_i}D_{CPU}^{V}]^T,
\end{equation}
\begin{equation} 
    {^u}{_i}D_{MEM} = [{^u}{_i}D_{MEM}^1,...,{^u}{_i}D_{MEM}^{V}]^T.
\end{equation}

Based on the Admission Control Policy, if a task requests unfulfillable amount of resources, then this task and the corresponding job should be rejected immediately.
If the soft-deadline is violated, the job will be redirected and rescheduled instead of being rejected permanently. 
Also, according to Service Level Agreement (SLA), one prerequisite should be met if a task ${^u}{_i}\varphi$ is allocated to server $\psi_n$ at time $t$: the requested CPU and memory in each VM type should be less than or equal to the amount of CPU and memory in that VM type on the server, i.e., 
\begin{equation} 
  {^u}{_i}D_{CPU}^v \leq R_{CPU}^v,\ {^u}{_i}D_{MEM}^v \leq R_{MEM}^v,\ \forall {v\in {^u}{_i}D_{VM}\subseteq V},
\end{equation}

\begin{figure*}
	\centering
	\includegraphics[width=\textwidth]{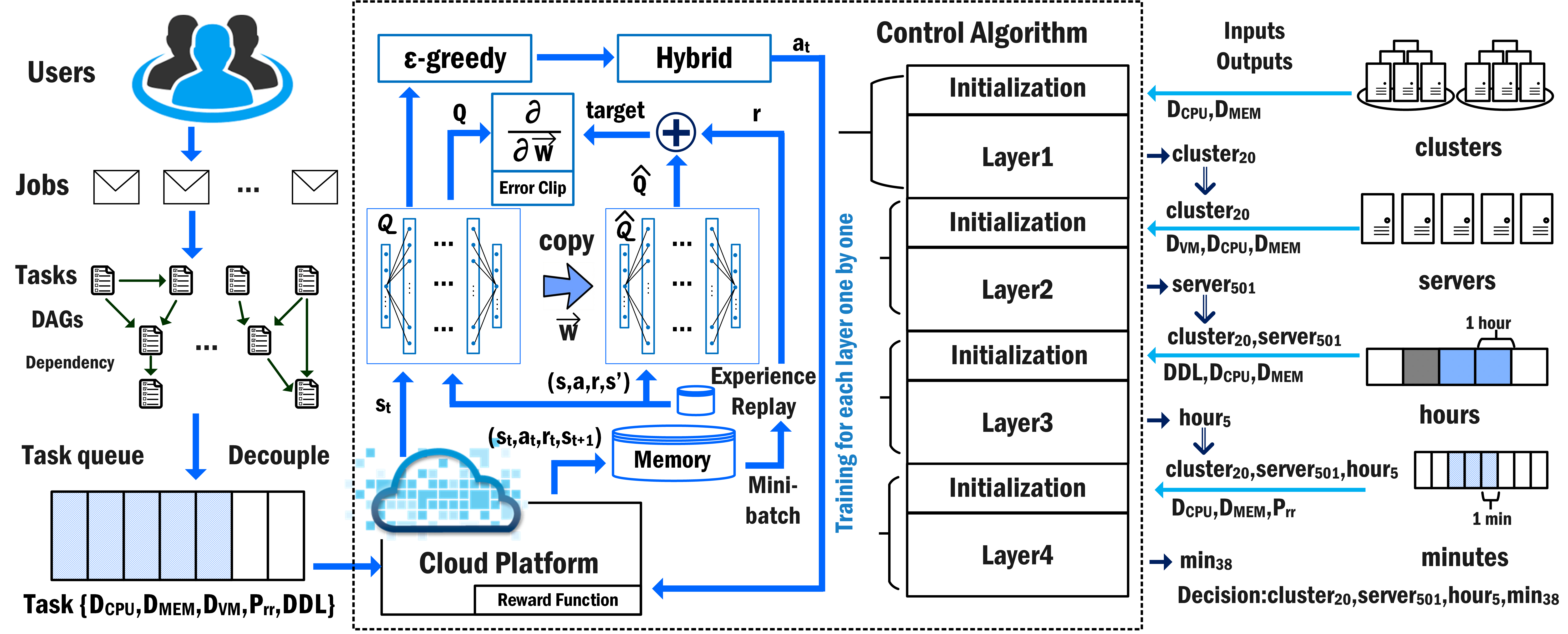}
	\vskip -3mm
	\caption{The structure of H$_2$O-Cloud: the details of workload model are described in Section \ref{workloadmodel}, those of the details of the hierarchical and hybrid framework are described in Section \ref{overviewofsystem}, and control algorithm is described in Section \ref{drlandcontrol}.}
	\label{fig:drlflow}
	\vskip -5mm
\end{figure*}

\subsection{Energy Consumption Model} \label{energyconsumptionmodel}
The total power of server $\psi_{n}$ at time $t$, $Pwr_{ttl}^n(t)$, is composed of static power $Pwr_{st}^{n}(t)$ and dynamic power $Pwr_{dy}^{n}(t)$, both of which are dependent on the CPU utilization rate $Ur^{n}(t)$ of server $\psi_{n}$ at time $t$, which is calculated as:
\begin{equation} 
  Ur^{n}(t) = \frac{\sum_{v\in V_n} R^v_{CPU}}{CPU_n}.
\end{equation}
$Pwr_{st}^{n}(t)$ is constant when $Ur^{n}(t)>0$ and zero otherwise.
$Pwr_{dy}^{n}(t)$ increases linearly as $Ur^{n}(t)$ increases linearly under the optimal utilization rate $Ur^{n}_{Opt}$, and increases non-linearly otherwise \cite{gao2013energy,li2017fast}. 
Additionally, different servers have different energy performances even under exactly identical utilization rate. 
In this paper, we adopt the dynamic power model from \cite{gao2013energy} and $Pwr_{dy}^{n}(t)$ is calculated as
\begin{equation}
  \begin{cases}
    Ur^{n}(t) \cdot a_{n}, & Ur^{n}(t)<Ur^{n}_{Opt}\\
    Ur^{n}_{Opt} \cdot a_{n}+\big(Ur^{n}(t)-Ur^{n}_{Opt}\big)^2 \cdot b_{n}, & Ur^{n}(t)\geq Ur^{n}_{Opt}.
  \end{cases}
\end{equation}
We use parameters $a_{n}$ and $b_{n}$ to capture the linear behavior and non-linear behavior of power increase of server $\alpha_{n}$, respectively. 
The total power used in all servers at time $t$ is $Pwr_{ttl}(t)= \sum_{n=1}^N Pwr^n_{ttl}(t)=\sum_{n=1}^N \big(Pwr^n_{st}(t)+Pwr^n_{dy}(t)\big)$.
The total energy consumption can be calculated as $Energy = \int_{0}^{T}Pwr_{ttl}(t)dt$, i.e., energy efficiency is $\frac{TotalCPU}{Energy}$, where $TotalCPU$ is the total amount of CPU requirement processed in time period $T$.

\subsection{Dynamic Pricing Model} \label{pricemodel}
In this paper, we consider a realistic non-flat price model $Price\big(t,Pwr\big)$ that is composed of a time-of-use-pricing (TOUP) component and a real-time pricing (RTP) component \cite{mohsenian2010optimal,abushnaf2015impact,li2017cts2m}.  
The TOUP is dependent on the time of the day and is usually higher in peak usage time periods than off-peak time, in order to incentivize users to shift loads towards off-peak periods. 
The RTP price increases as the total energy usage grows.
The energy cost of server $\psi_{n}$ at time $t$ is $Price(t,Pwr_{ttl}^n(t))$.
The total energy cost of all servers is:
\begin{equation} \label{eq:totalcost}
	Cost = \Sigma_{t=1}^T Price\big(t,Pwr_{ttl}(t)\big).
\end{equation}
Then energy cost efficiency can be calculated as $\frac{TotalCPU}{Cost}$, where $TotalCPU$ is the total amount of CPU requirement processed in time period $T$.

\subsection{Problem Formulation} 
In this work, the cloud energy cost reduction problem is formulated as:\\
\noindent\textbf{Given} the user workload model, cloud platform model, energy consumption model, and dynamic pricing model.

\noindent\textbf{Find} resource allocation and task scheduling on-the-fly.

\noindent\textbf{Maximize} energy cost efficiency.

\noindent\textbf{Subject to:}
\begin{equation}
\Sigma_{v\in V_n} {R}_{CPU}^v(t) \leq CPU_n, \forall t, \forall n
\end{equation}
\begin{equation}
\Sigma_{v\in V_n} {R}_{MEM}^v(t) \leq MEM_n, \forall t, \forall n
\end{equation}
\begin{equation} 
{^u}{_i}D_{CPU}^v \leq R_{CPU}^v(t), \forall {{^u}{_i}\varphi}, \forall v, \forall t, \forall n
\end{equation}
\begin{equation}
{^u}{_i}D_{MEM}^v \leq R_{MEM}^v(t), \forall {{^u}{_i}\varphi}, \forall v, \forall t, \forall n
\end{equation}
$$and\ task\ dependency\ requirements.$$

\section{H$_2$O-Cloud: The Resource Management and Task Scheduling Framework} \label{overviewofsystem}

Making decisions under a comprehensive system model and in one step may offer high performance as the overall environment and user request information are considered altogether. 
That however would significantly increase the runtime as the state space and possibly the action space are large; which in turn eliminates the possibility of real-time effectiveness and hence impacts its feasibility.

In this section, we present H$_2$O-Cloud: a \underline{H}ierarchical and \underline{H}ybrid \underline{O}nline \underline{Cloud} resource management and task scheduling framework based on DRL.
H$_2$O-Cloud utilizes cloud resources efficiently and is aware of QoS.
As shown in Figure \ref{fig:drlflow}, four decisions are made in the proposed multi-layer system. More specifically, H$_2$O-Cloud provisions cloud resources in clusters and servers in the first two layers, and schedules tasks into appropriate hours and specific minutes in the last two layers.
Training of Deep Q-Network (DQN) systems is done on-the-fly, and performance is guaranteed by fast convergence of our algorithm described in Section \ref{training}.
By using DRL-based hybrid Algorithm \ref{Algorithm: DeepQlearning} we have proposed in this work, resource efficiency and QoS are improved.

\subsection{Multi-Layer Resource Management and Task Scheduling Framework} \label{thesystem}
A CSP with tens of thousands of DCs faces a complicated decision-making challenge as it receives millions of user requests a day, and has to figure out the following decisions to run a certain incoming task:
1) which server cluster; 2) which server in that cluster; 3) which hour of a day to reduce electric bill; 4) which minute(s) in that hour to ensure on-the-fly response and avoid congestion; and 5) how to manage the deadline and priority, e.g., reschedule the task if its soft-deadline cannot be met.
All of these decisions need to be made in a short time (less than one minute) when an incoming task has arrived. 

In this work, we consider a CSP with warehouse-scale servers.
Tens of thousands of servers are clustered into tens of server clusters.
In order to make decisions more efficiently, and also increase the performance of the DQN of our system, our framework first decides which cluster should be used and then specifies the exact server. 
In addition, according to realistic price model and quality of service awareness, coarse-grained scheduling (hour control) and fine-grained scheduling (minute control) of tasks should be done separately to meet the requirement of efficiency and effectiveness. 

\subsubsection{Layer 1 - Cluster-Level Control}
The inputs of this layer, $D_{CPU}$ and $D_{MEM}$, as shown in Figure \ref{fig:drlflow} are used in the \emph{Admission Control} to filter jobs with tasks demanding unfulfillable resources. 
After admission control, one task could be handled by any cluster candidate in the warehouse-scale CSP. In order to reduce energy cost and latency of transmission between clusters, tasks within one job will be allocated into the same cluster. 
And the first layer decides which cluster to allocate the job.
With proper training, this DQN can learn the rules of assigning new jobs to a cluster such that overall the $Ur$ is optimized.

\subsubsection{Layer 2 - Server-Level Control}
One cluster may consist of tens of thousands of servers, and each server supports several different kinds of virtual machine configurations.
Server-level control allocates incoming tasks into a subset of servers which have the required machine configurations to support these tasks.
This is the reason to include the VM type $D_{VM}$ in the inputs of this layer. 
The second layer then decides which server to host the task. 
With the deep structure in the DQN, a rule is learned to select a server with medium $Ur$ (e.g. $40\%-60\%$) to host incoming tasks as it is closer to the optimal utilization rate.
The admission control in this layer is done similarly as cluster-level.

\subsubsection{Layer 3 - Hour Scheduling}
The realistic electric price model comes into play in this layer.
Incoming tasks are highly likely to be allocated into hours with lower electric prices if $Ur$ is in an acceptable (efficient) range.
In other words, hour scheduling is congestion-aware, cost-aware, and resource-efficient.
All incoming tasks come with soft-deadlines designated by end-users, and the proposed system will assign a penalty if any soft-deadlines is violated.
Therefore, hour scheduling is also deadline-aware and QoS-aware.
The admission control in this layer is done similarly to that of the cluster-level.

\subsubsection{Layer 4 - Minute Scheduling}
The proposed system schedules tasks into specific server and exact minute(s) as the requirement of online scheduling.
Tasks with higher priorities will be executed earlier, which fulfills the priority-aware requirement.
This last layer is QoS-aware, as it provides feedback to the scheduler when priority requirement is met or violated.
The admission control is the same as the one in the cluster-level.

\subsection{Deep Q-Learning Background}
Deep Q-learning combines reinforcement learning with deep neural network, which is proposed by Google DeepMind \cite{deepmind} and used in playing Atari games \cite{mnih2013playing,mnih2015human}.
In an environment, a reinforcement learning (RL) agent makes a sequence of actions ($a$), based on the observations of environment and rewards ($r$) of actions. 
At each time-step the agent selects a valid action from a set of available actions, which is called the action space.
By taking the action, the agent changes the internal state of the environment, which may not be accessible to the agent, instead the agent only obtains some observation of the environment.
The agent can also get some rewards in addition to observation after taking the action.
Therefore, the inputs of RL algorithm are the sequence of previous actions and observations of the environment, and the algorithm learns strategies to maximize future rewards depending on these sequences ($s$).

As we mentioned before, the goal of RL agent is to seek maximum future rewards by selecting actions. 
The future rewards are discounted by a factor of $\gamma$ per time-step, so the future discounted reward at time $t$ is defined as $R_t = \sum_{t'=t}^{T}\gamma^{(t'-t)}\cdot r_{t'}$.
Then the maximum expected return achievable by following any policy $\pi$ after seeing sequence $s$ and taking action $a$ is defined as the optimal action-value function $Q^*(s,a)=max_\pi E[R_t|s_t=s,a_t=a,\pi]$. 
This optimal action value function obeys the identity known as the Bellman equation:
\begin{equation}\label{eq: bellman}
Q^*(s,a) = E_s'[r+\gamma max_{a'} Q^*(s',a')|s,a]
\end{equation}

It is common to use a function approximator to estimate the optimal action-value function:
$Q(s,a;\theta) \approx Q^*(s,a)$.
Other than a linear function approximator, a nonlinear function approximator such as a neural network can also be used.
A neural network function approximator with weights $\theta$ is defined as a Q-network \cite{mnih2015human}.

\subsection{Deep Q-Learning in Resource Management System} \label{state_action_reward}
For a warehouse-scale CSP, the state space and action space are enormous if tasks are allocated directly into specific server and minute(s).
In order to overcome this challenge, we propose the multi-layer structure to decouple the huge decision spaces (i.e., number of clusters $\times$ number of servers $\times$ number of hours $\times$ number of minutes in one hour).
Each layer of H$_2$O-Cloud uses deep Q-learning algorithm with experience play (shown in Algorithm \ref{Algorithm: DeepQlearning}) which is suitable for solving complicated control and decision-making problems with a large state space but a finite action space. 
Also, the four-layer framework divides the original one single action space with an enormous size into four smaller subsets with different input features. 

The algorithm selects and executes decisions according to a deep Q-learning based \textit{hybrid} algorithm, i.e., if the $\epsilon$-greedy policy generates a legal action, then the action will be executed and stored into memory. Otherwise the Round-Robin algorithm will be used to generate the action.
The rationale behind this is that DRL technology sometimes generates illegal actions and with the same input, the decision-making neural network generates same output before next learning.
Therefore, the background Round-Robin algorithm can help to avoid poor decisions in such cases.

\begin{algorithm}[t]\footnotesize
 \label{Algorithm: DeepQlearning}
	\SetAlgoLined
	\DontPrintSemicolon
	Initialize replay memory to size $\Delta$ \\
	Initialize action-value function $Q$ with random weights $\theta$\\
    Initialize target action-value function $\hat{Q}$ with weights $\theta' = \theta$\\
	\For{$episode = 1, E$}{
    	Run $env.reset(layer)$ to set cloud environment to initial state\\
        Initialize sequence $s_1 = \{x_1\}$\\
		\For{$t = 1, T$}{
        	With probability $\epsilon$ choose a random action $a_t$\\
            otherwise choose $a_t = argmax_aQ(s_t,a;\theta)$\\
            Execute action $a_t$ and observe next observation $x_{t+1}$, reward $r_t=R(layer)$, reject signal $reject$, and valid action number $option$ ($option \in [0,Dimension(Action\ Space)]$)\\
            \While {$reject = 1$ and $option > 1$}{
	            $round\_robin()$, get action $a_t'$\\
                \If {$a_t \neq a_t'$}{
                    Replace $a_t$ with $a_t'$,\\
                    Execute action $a_t$ and observe next observation $x_{t+1}$, reward $r_t$, reject signal $reject$, and valid action number $option$ \\
               	}
            }
            Set $s_{t+1} = s_t,a_t,x_{t+1}$ \\
            Store transition $(s_{t+1},a_t,r_t,s_t)$ in memory\\
            Store decision and reject signal\\
            Sample random mini-batch of transitions $(s_{j+1},a_j,r_j,s_j)$ from memory\\
            $ target_j= \begin{cases} 
r_j,\text{if episode terminates at step j+1 }\\
r_j+\gamma max_{a'}\hat{Q}(s_{j+1},a';\theta '),\text{otherwise}
\end{cases}$\\
		Perform a gradient descent step on $\big(target_j - Q(s_j,a_j;\theta)\big)^2$ \\
        Every $Y$ steps, train evaluation network, decrease $\epsilon$\\
        Every $\zeta$ steps, copy $Q$ to $\hat{Q}$\\
        }
	}
    \KwRet All actions $a_t$, all reject signals
	\caption{Deep Q-Learning for H$_2$O-Cloud With Experience Replay}
\end{algorithm}

\subsubsection{Action Space} 
A RL agent interacts with its environment and selects an action from a set of available actions based on observation.
In our case the proposed framework receives an incoming task, and selects a server to execute this task at some time (i.e., action), based on the information retrieved from the task and data center environment, such as requested resources of task, available resource of servers, task priority and deadline designated by end user, and historical decisions made by agent, etc.
For layer 1 to layer 4, action spaces are defined as: available clusters, available servers, available hours, and available minutes, respectively. 

\subsubsection{State Space}
The agent observes current status of the data center and incoming task, and the optimal action is determined based on current observation $x$.
Current server observation describes available resources such as CPU and memory of requested VMs on server, whereas the observation of task ${^u}{_i}\varphi$ (i.e., task $i$ in $job_u$) is the parameter tuple $\{{^u}{_i}D_{CPU},{^u}{_i}D_{MEM},{^u}{_i}D_{VM},{^u}{_i}Prr,{^u}{_i}DDL\}$ presented in Section \ref{workloadmodel}.
Therefore, state $s_t$ at time $t$ is a sequence of actions and observations, i.e., $s_t = x_1,a_1,x_2,a_2,...,a_{t-1},x_t$ \cite{mnih2015human}, and is input to the proposed deep Q-learning-based system. 
After taking an action, the RL agent receives an immediate reward from the environment, and transits into next state.
Instead of using immediate reward as the value of state-action pair, Q-learning agent updates the Q value as Equation \ref{eq:Qvalueupdate} obeys Bellman equation \ref{eq: bellman}, which is a weighted sum of the expected values of the rewards of all future steps starting from the current state, in its value iteration.
\begin{multline}\label{eq:Qvalueupdate}
Q_{t+1}(s_t,a_t) = Q_t(s_t,a_t) +\\ 
 \eta\big(r_{t+1} + \gamma \max\limits_{a_{t+1}}Q_t(s_{t+1},a_{t+1})-Q_t(s_t,a_t)\big)
\end{multline}
with learning rate $\eta \in (0,1]$ and discount factor $\gamma \in [0,1]$. 

\subsubsection{Reward Function} \label{rewardfunctiondef}
The goal of H$_2$O-Cloud is to maximize cost efficiency while maintaining QoS by taking a sequence of actions.
Therefore, reward functions in four layers are modified differently to achieve different goals for training purposes.
After running lots of experiments to verify the reward function hypothesis, this is the best combination we've found in our experiments.

In Layer 1, reward function $R(layer = 1)$ is defined as:
\begin{equation}
  \begin{cases}
    +1, & 0\%\leq Ur(t)<45\%\\
    -2, & 50\%< Ur(t)\\
    -1, & otherwise
  \end{cases}
\end{equation}
In Layer 2, reward function $R(layer = 2)$ is defined as:
\begin{equation}
  \begin{cases}
    +1, & 20\%\leq Ur(t)<80\%\\
    -2, & 100\%< Ur(t)\\
    -1, & otherwise
  \end{cases}
\end{equation}
In Layer 3, reward function $R(layer = 3)$ is defined as:
\begin{equation}
  \begin{cases}
    -2, & 100\%< Ur(t)\\
    -Price(t,Pwr_{ttl}^n(t)) & Price(t,Pwr_{ttl}^n(t))<0.3\\
    -Price(t,Pwr_{ttl}^n(t))*4 & otherwise
  \end{cases}
\end{equation}
In Layer 4, reward function $R(layer = 4)$ is defined as:
\begin{equation}
  \begin{cases}
    -2, & 100\%< Ur(t)\\
    -1, & 80\%< Ur(t)\ or\ Ur(t)<20\% \\
    +2, & 60\%< Ur(t) \leq 80\%\\
    +1, & otherwise
  \end{cases}
\end{equation}
If the scheduled minute $\in [{^u}{_i}Prr-1,{^u}{_i}Prr+1]$:
\begin {equation} \label{layer4reward}
R(layer=4) = R(layer=4)*2.5
\end{equation}

The modification of reward function parameters is performed empirically based on our experimental results, and according to the following two principles: (i) Parameter modifications in the current layer highly impacts the performance of the following layers. We therefore manually select cases with potentially positive impacts, in order to minimize the chance of cases with negative effects on the following layers.
(ii) The preferred range of $Ur(t)$ for reward function of each layer is selected according to practical limits and revised based on principle (i). For example, in Layer 4, the reward function takes a maximum reward value of $+2$ in case the utilization rate of a server in the exact minute ranges in $(60\%, 80\%)$. Higher utilization rates provide higher resource effectiveness, and naturally leads to a positive reward feedback. However, during the training process of DQN, if the maximum reward is assigned based on utilization rate in the range $(80\%,100\%)$, the DQN will learn to assign as many tasks as possible to that range. That may result in high rejection rates and impact the overall performance of the framework. Hence, the maximum reward should be assigned to range lower than $(80\%,100\%)$ to encourage scheduling and prevent rejection at the same time.

\section{H$_2$O-Cloud: The Deep Q-learning and System Control Algorithm} \label{drlandcontrol}

For a warehouse-scale CSP, the key target is to schedule a large volume of incoming tasks rapidly and effectively.
The decomposition strategy of H$_2$O-Cloud avoids a large DQN and guarantees on-the-fly decision-making capacity, as training four smaller DQNs converges much faster than training a huge one. 
In addition, there are four types of decision (i.e., those related to cluster, server, hour and minute(s)) that need to be made in the whole scheduling process. Different feature will cause different effects to DQN. A combination of features may lead to slower convergence of algorithm and degradation of decision quality. 
Finding the optimal reward function for a DQN is a difficult problem, especially when mixing different features in the decision-making process.
Therefore, dedicating one DQN to each decision type not only provides reasonable runtime, but also further guarantees final scheduling decision quality.

Scheduling order is designed as cluster, server, hour, and minute for the following reasons:
(i) It is obvious that the server farm and hour should be designated before choosing the exact server and minute(s).
(ii) Scheduling efficiency will be higher if we first determine server farm and server, then select hour and minute(s). 
For example, assume an incoming task requests unfulfillable amount of resource, and it should be rejected at server layer by Admission Control.
If we use the hour, minute(s), server farm, and server scheduling order, the task will be rejected at the last layer and the decision-making steps of the previous three layers are wasted.
However, if we use the server farm, server, hour, and minute(s) scheduling order, the task will be rejected at the second layer, which is more efficient in terms of runtime. 
(iii) First choosing server, then selecting an optimal hour, yields a reward function (described in Section \ref{rewardfunctiondef}) of DQN that is based on energy cost of one server. 
However, if we decide which hour to run an incoming task first, the reward function will be based on total energy cost of all servers. 
It will be hard to guarantee that each server is working in its optimal RUtR and energy cost range, although both scheduling orders are valid in task scheduling problem.

\begin{algorithm}[b]\scriptsize
\label{Algorithm1}
	\SetAlgoLined
	\DontPrintSemicolon
	Initialize realistic price model $Price(t,Pwr)$\\
	Initialize environment as $env$ \\
    Layer I: Initialize $DQN(layer = 1)$ for cluster control\\
    Run $DQN()$, store user request allocation $A_{c}$ and reject signals $Reject_{c}$\\
	Layer II: Initialize $DQN(layer = 2)$ for server control\\
	\For{all cluster}{
        Run $DQN(cluster)$, store $A_{s}$ and  $Reject_{s}$
	}
    Layer III: Initialize $DQN(layer = 3)$ for hour control\\
	\For{all server}{
        Run $DQN(cluster,server)$, store $A_{t}$ and $Reject_{t}$
	}
    Layer IV: Initialize $DQN(layer = 4)$ for minute control\\
	\For{all hour}{
        Run $DQN(cluster,server,hour)$, store $A_{m}$ and $Reject_{m}$
	}
    Calculate final user request allocation matrix, i.e., $Ur$ for every server in every minute according to $A_{c}$, $A_{s}$, $A_{t}$, $A_{m}$, $Reject_{c}$, $Reject_{s}$, $Reject_{t}$ and $Reject_{m}$\\
    Calculate final energy consumption $Energy$, electric bill $Cost$, energy efficiency and cost efficiency\\
	\caption{Control Algorithm for H$_2$O-Cloud}
	\label{Algorithm:control}
\end{algorithm}

\subsection{System Control Algorithm}
System control algorithm in this work is shown in Algorithm \ref{Algorithm1}, which implements the hierarchical structure described in Section \ref{thesystem}. 
After initialization of the price model and environment of DRL, that includes the user workload model and cloud platform model, our four-layer decision-making processes are executed as shown in Algorithm \ref{Algorithm1}.
$DQN()$ is the proposed Algorithm \ref{Algorithm: DeepQlearning}, which is modified for each layer, as described before. 
Decisions of tasks are recorded for the later calculation of energy-efficiency and cost-efficiency, etc.

Each layer takes output of preceding layers as input, and uses Algorithm \ref{Algorithm: DeepQlearning} as the decision-making agent.
The training method is used in all four layers, so there are four such training procedures for one batch of data.
All four layers in H$_2$O-Cloud utilize the same DQN hidden layer structure and deep Q-learning parameters, such as the learning rate, discount rate, etc. 
Algorithm \ref{Algorithm: DeepQlearning} converges fast in all four layers, which guarantees the good performance, and shows robustness and portability of our algorithm.

Figure \ref{fig:drlflow} shows the structure of our H$_2$O-Cloud, with an example.
For layer 1, after $DQN()$ is initialized, it takes available clusters and incoming tasks as input, and makes decision according to Algorithm \ref{Algorithm: DeepQlearning}. 
Action space in layer 1 of H$_2$O-Cloud agent is the available clusters. 
Reward function is defined based on $Ur$ as described in Section \ref{rewardfunctiondef}. 
For instance, after taking an action (e.g., $cluster_{20}$ as shown in Figure \ref{fig:drlflow}), a positive reward will be received when $Ur$ of $cluster_{20}$ is lower than $45\%$.

Cluster-level control is done for incoming tasks in Layer 1. Then the server-level control will proceed in Layer II.
Initialization of layer 2 takes output of layer 1, for example $cluster_{20}$, and information of task $\{D_{VM}, D_{CPU}, D_{MEM}\}$ as inputs.
Each server in $cluster_{20}$ may hold a subset of VMs as described in Section \ref{cloud_platform_model}, therefore the action space of our H$_2$O-Cloud agent in layer 2 is the available servers in the cluster, e.g., $cluster_{20}$. 

Output of layer 2 (e.g., $server_{501}$), layer 1 (e.g., $cluster_{20}$), and information of incoming task $\{DDL, D_{CPU}, D_{MEM}\}$ will then be the input of layer 3. 
Next, layer 3 schedules the task into an hour (e.g. $hour_{5}$ in our example).
The last decision, i.e., exactly which minute(s) to spend for the task execution, is made by the final layer, i.e., layer 4.
This completes the decision procedures of the multi-layer part of our H$_2$O-Cloud framework. 
For example, the complete decision in Figure \ref{fig:drlflow}  is represented as $cluster_{20},server_{501},hour_{5},min_{38}$. 
\subsection{Training for Deep Q-Learning Algorithm in H$_2$O-Cloud} \label{training}
\subsubsection{Hybridity}
DRL agents may produce invalid actions while training on-the-go, such as actions violating the soft-deadline that is designated by end-user, or actions that make resource utilization of server explode, etc.
In addition, DRL tends to make the same decision when current state stays the same, i.e., it will keep scheduling an incoming task into same place. 
Therefore, instead of waiting for the DRL agents to make a valid decision for current task by the $\epsilon$-greedy mechanism, our hybrid method can provide a valuable action more efficiently.
In other words, the hybrid method is a backup plan which will be triggered if DRL agent cannot handle the incoming task efficiently.
Furthermore, incoming tasks must be scheduled in a short time after they have arrived. This is rather significant for a warehouse-scale CSP to provision resources and schedule tasks on-the-fly. 
Therefore, Algorithm \ref{Algorithm: DeepQlearning} employs Round-Robin, a fast scheduling mechanism, in a hybrid fashion along with the DQN  systems to schedule tasks into valid positions rapidly.

\subsubsection{Experience Replay and Target Network}
Q-learning updates are done in mini-batches of experience, which are selected randomly from the memory of stored transitions. 
Mini-batches training breaks the correlation of learning data and reduces the variance of update, which provides learning procedure with a higher efficiency.
Target network is a separate neural network with the same structure as the evaluation network, which is used in DQN to generate target Q value in the Q-learning update.
By introducing a delay between the time an update affects Q and the time an update affects the target, divergence and oscillations are eliminated \cite{mnih2015human}. 

\subsubsection{Error Clipping}
In this work, we also use error clipping \cite{mnih2015human} to further improve the stability of Q-learning algorithm. 
By clipping the error term from the update $\big(target_j - Q(s_j,a_j;\theta)\big)^2$ into $[-1,1]$ interval, convergence of Algorithm \ref{Algorithm: DeepQlearning} is further guaranteed.

\section{Experimental Results} \label{experiment}
\subsection{Experiment Setup} 
\subsubsection{Baselines} The following three baselines are used to evaluate the efficiency of H$_2$O-Cloud:
\begin{itemize}
\item \textbf{Round-Robin (RR):} RR is commonly used as a baseline in cloud resource management \cite{katyal2014comparative, pradhan2016modified}. 
The CSP assigns each task in a circular order. If the current assignment violates SLA, the scheduler will try the following options until non violation. If no possible assignment, the task will be rejected as unfulfillable.
In the experiment setup of this work, we have chosen RR as a baseline for its scalability and reasonable runtime.
\item \textbf{Hierarchical-DRL (HDRL):} This baseline is a modified, non-hybrid version of our H$_2$O-Cloud, i.e., it depends on the hierarchical DRL only, without hybridity.
This baseline helps us evaluate the improvements contributed by hybridity on top of the hierarchical DRL.
\item \textbf{DRL-Cloud:} This baseline is proposed in \cite{cheng2018drl}, which uses a DRL-based two-stage Resource Provisioning and Task Scheduling (RP-TS) processor to schedule tasks into hours without fine-grained scheduling and hard-deadlines only. 
\end{itemize}

Three user workload scenarios (i.e., user-server relations) are used (Figure \ref{fig:user_request}).
They are as follows: $\#user \ll \#server$, $\#user < \#server$, $\#user \simeq \#server$, which considers all the possible realistic scenarios.
We adopt the realistic price data from \cite{electricalprice,faruqui2010ethics} to calculate the energy cost after scheduling, where it is assumed that the utility company announces the pricing $24$ hours in advance.
Google cluster-usage traces with 12.5k servers and timing information \cite{googletrace} are used in this work.

\subsubsection{Indicators} Comparisons are based on five indicators:
\begin{itemize}
\item \textbf{Energy Cost Efficiency (ECE):} Units of CPU that can be executed by using one unit of energy cost. 
\item \textbf{Energy Efficiency (EE):} Units of CPU that can be executed by using one unit of energy. 
\item \textbf{Turn-off Rate (TFR):} Servers will be turned off when no task is scheduled into it in one hour. 
TFR is introduced to visualize the sparse usage of servers.
\item \textbf{Util-opt Rate (UOR):} Util-opt rate indicates how may working servers are running in their optimal status. Compared to TFR, which shows sparse usage of servers, UOR shows resource utilization status of working servers as described in Section \ref{intro}. 
\item \textbf{ddl-Violation Rate (ddlVR):} Soft-deadline violation rate.
 \item \textbf{Reward Rate:} A reward will be given to the scheduler in training process when priority requirement is fulfilled. Reward Rate indicates how often the scheduler receives the priority fulfillment reward, i.e., it represents how often the user-designated task priority requirement can be fulfilled.
\end{itemize}

The data details for the low, medium and high variance scenarios, shown in Figure \ref{fig:user_request}, Figure \ref{fig:mediumvar}, Figure \ref{fig:vsdrl}, Table \ref{tbl:lowhighvar} and Table \ref{tbl:vsdrl}, appear below. They are all extracted from Google trace real traffic: 
\begin{itemize}
\item \textbf{Numbers of Tasks:} $77776$, $154001$ and $265865$.
\item \textbf{Requested CPU Units:} $2610.98$, $5776.82$, and $9488.66$.
\item \textbf{Variances of CPU:} $1621.64$, $16055.58$, and $42462.99$.
\item \textbf{Variances of memory:} $479.59$, $13543.92$, and $23996.92$.
\end{itemize}

\subsubsection{Experiment Scale} Comparisons are performed based on the following three scales:
\begin{itemize}
\item \textbf{Small-Scale:} $600$ servers in total, clustered into $2$ server clusters, corresponding to $\#user \simeq \#server$ user-server relation.
\item \textbf{Medium-Scale:} $1,080$ servers clustered into $3$ server clusters, corresponding to $\#user < \#server$.
\item \textbf{Large-Scale:} $12,500$ servers clustered into $5$ server clusters, corresponding to $\#user \ll \#server$.
\end{itemize}

DQN structure in both HDRL and H$_2$O-Cloud uses a two-hidden-layer neural network with ten neurons in each hidden layer.
According to our experiments, hierarchical strategy of HDRL and H$_2$O-Cloud makes it less sensitive to DQN structure changes, as long as the number of hidden layers and the number of neurons lie in the range of $[2,5]$ and $[5,50]$, respectively.
For deep Q-learning, learning rate $\eta = 0.1$, discount factor $\gamma = 0.9$, and $\epsilon$ is decreased from $0.9$ by $0.05$ in each learning iteration, memory size $\Delta=500$, replace target network step $\zeta= 300$, train evaluation network step $Y=20,5,10,10$ in Layer 1 to 4, respectively.
All parameters are initially chosen from a commonly used range \cite{mnih2015human, strehl2006pac} and tuned based on the experiments in order to provide relatively optimal performances.

To show the effectiveness of our framework more clearly, the following comparisons are presented in the following: 
Our HDRL and H$_2$O-Cloud vs. RR (Section \ref{exp:HHvsRR}), HDRL and H$_2$O-Cloud vs. DRL-Cloud (Section \ref{exp:HHvsDRL}), and H$_2$O-Cloud vs. HDRL (Section \ref{exp:HvsH}).

\begin{table*}[!h]
	\centering
	\vskip -0.5cm
	\caption{Comparison of H$_2$O-Cloud and HDRL with RR under Low Variance, and High Variance Scenarios in Small, Medium, and Large Scales}\label{tbl:lowhighvar}
	\resizebox{1\textwidth}{!}{
		\begin{tabular}{c|c|ccc cc|ccc cc   }
			\hline
            \multicolumn{2}{c}{} & \multicolumn{5}{|c|}{Low Variance (77776 tasks, 2610.98 cpu in 1.39 hour trace)} & \multicolumn{5}{c}{High Variance (265865 tasks, 9488.66 cpu in 1.39 hour trace)}\\\hline
            Scale & Approach & Norm. ECE & Norm. EE & UOR & TFR & ddlVR & Norm. ECE & Norm. EE & UOR & TFR & ddlVR \\\hline
  \multirow{5}{*}{Large}
  			& RR & 0.23 & 0.80 & 0.00\% & 0.00\% & 22.80\% 
            & 0.21& 0.83& 0.00\%& 0.00\%& 22.91\%\\\cline{2-12}
            & HDRL & 0.88 & 0.99 & 11.46\% & 92.36\% & 10.19\% 
            & 0.99 &1.00 &4.00\% &89.31\% & 16.28\%\\
            & improve & 283.18\% & 22.77\% & $\infty$ & $\infty$ & 55.31\% 
            & {371.43}\%& 20.80\%& $\infty$ & $\infty$ & 28.94\% \\\cline{2-12}
            & H$_2$O-Cloud & 1.00 & 1.00 & 7.42\% & 92.60\% & 8.23\% 
            & 1.00& 1.00& 4.81\% & 87.14\% &13.89\% \\
            & improve & 333.57\% & 24.34\% & $\infty$ & $\infty$ & 63.90\% 
            & 376.19\% & 21.18\% & $\infty$ & $\infty$ & 39.38\% \\\hline
  \multirow{5}{*}{Medium}
  			& RR & 0.22 & 0.65 & 0.00\% & 0.00\% & 22.92\% 
            & 0.21& 0.60& 0.00\%&0.00\% &22.92\% \\\cline{2-12}
            & HDRL & 1.00 & 1.00 & 25.21\% & 53.63\% & 7.41\% 
            & 0.88& 0.93&15.00\% &37.36\% &8.68\% \\
            & improve & 359.81\% & 53.20\% & $\infty$ & $\infty$ & 67.67\% 
            & 319.05\%& 54.77\% & $\infty$ & $\infty$ & 62.48\% \\\cline{2-12}
            & H$_2$O-Cloud & 0.97 & 0.99 & 19.13\% & 45.90\% & 7.01\% 
            & 1.00 & 1.00 & 13.44\% & 28.70\% & 6.02\% \\
            & improve & 346.19\% & 52.39\% & $\infty$ & $\infty$ & 69.42\% 
            & 376.19\% & 67.15\% & $\infty$ & $\infty$ & 73.73\% \\\hline            
  \multirow{5}{*}{Small}
  			& RR & 0.23 & 0.57 & 0.00\% & 0.00\% & 22.83\% 
            & 0.25 & 0.59 & 0.00\% & 0.00\% & 22.91\% \\\cline{2-12}
            & HDRL & 1.00 & 1.00 & 19.78\% & 22.04\% & 6.55\% 
            & 0.74 & 0.84 & 14.00\%  & 7.92\% & 10.23\% \\
            & improve & 331.41\% & {75.39}\% & $\infty$ & $\infty$ & 71.31\% 
            & 196.00\%&43.07\% & $\infty$ & $\infty$ & 55.35\%\\\cline{2-12}
            & H$_2$O-Cloud & 0.78 & 0.66 & 24.17\% & 40.70\% & 14.26\% 
            & 1.00& 1.00& 16.93\% & 13.38\% & 6.05\% \\
            & improve & 238.51\% & 15.22\% & $\infty$ & $\infty$ & 37.54\% 
            & 300.00\% & {70.13}\% & $\infty$ & $\infty$ & 73.81\% \\\hline    			
		\end{tabular}
	}
\end{table*}

\subsection{H$_2$O-Cloud and HDRL vs. RR} \label{exp:HHvsRR}
\begin{figure}[!h]
	\centering
	\includegraphics[width=0.95\columnwidth]{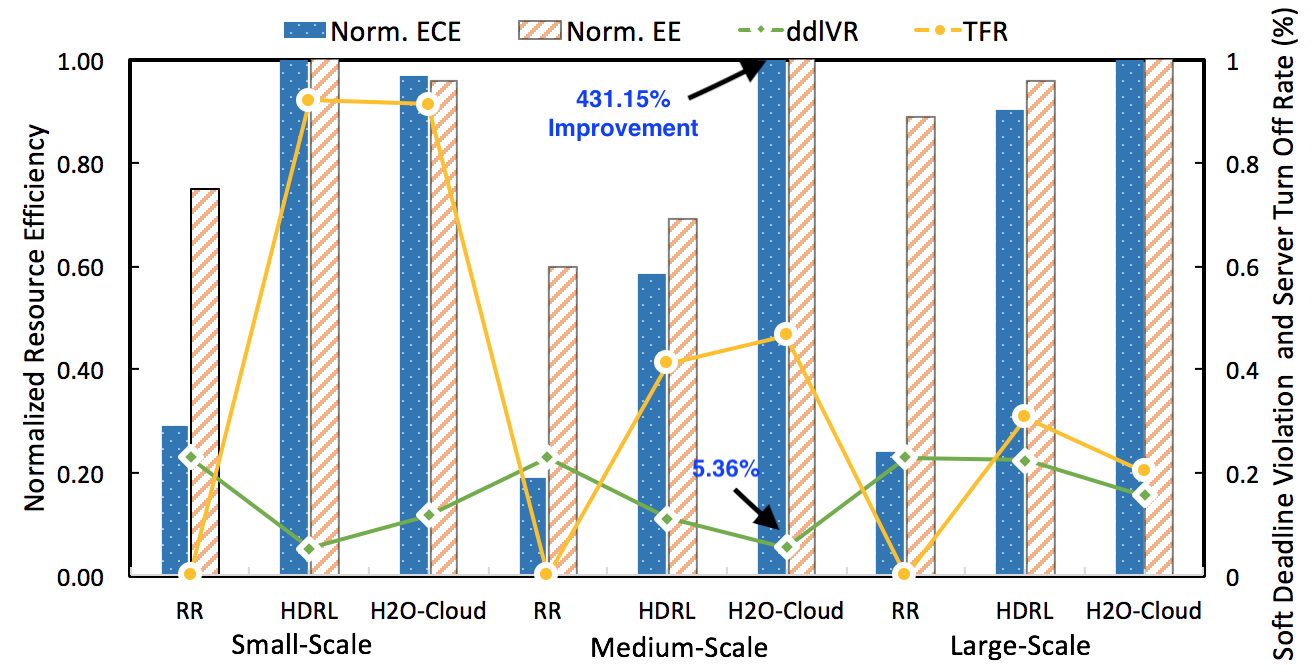}
	\vskip -4mm
	\caption{Comparison of H$_2$O-Cloud, HDRL and RR under medium variance scenario in small, medium and large scales.}
	\label{fig:mediumvar}
	\vskip -2mm
\end{figure}

Comparisons of H$_2$O-Cloud and HDRL with baseline RR in low variance and high variance scenarios are shown in Table \ref{tbl:lowhighvar}, with RR as the reference (base). The normalization process uses the highest value in each column as the reference. Comparison in the medium variance scenario is shown in Figure \ref{fig:mediumvar}. Compared to RR, H$_2$O-Cloud and HDRL significantly outperform in terms of ECE, EE, TFR, UOR and ddlVR, which shows that H$_2$O-Cloud and HDRL achieve higher energy and cost efficiency while maintaining QoS.

Based on admission control policy, if a task requests unfulfillable amount of resources, then this task should be rejected immediately. 
All the baselines and H$_2$O-Cloud follow the same admission control policy and the implementation codes are strictly the same. 
RR completes admission control in one step, while HDRL and H$_2$O-Cloud perform it hierarchically. These two approaches are fundamentally very similar.
Therefore, the results are not biased by the admission control. 
Same admission policy means that the rejection rates of difference policies are very similar. 
In our experiment setup, RR would not reject tasks as they do not turn off servers. The rejection rates of both HDRL and H$_2$O-Cloud range between $0\%$ and $0.1\%$ in our various setup cases.
RR provides zero task rejection and results in much higher energy consumption in return. HDRL and H$_2$O-Cloud balance between energy cost efficiency and task rejection rate by scheduling tasks and turning off servers intelligently while maintaining QoS. It should be noted that HDRL and H$_2$O-Cloud can be modified to achieve zero task rejection while outperforming RR in terms of energy efficiency if we put rejection rate as top priority.

As we can see in Table \ref{tbl:lowhighvar}, 
H$_2$O-Cloud and HDRL consistently outperform RR in all scenarios in terms of normalized ECE, normalized EE, UOR, TFR and ddlVR. H$_2$O-Cloud and HDRL result in significant improvement because they are both aware of historical decisions and current environment status.
In addition, utilization of user request information such as task priority and task deadline lead to low soft-deadline violation, which improves quality of service.
High turn-off rate shows the sparse usage of DCs, which further improves resource efficiency in working servers.

According to our experiments in Table \ref{tbl:lowhighvar} and Figure \ref{fig:mediumvar}, H$_2$O-Cloud and HDRL achieve up to $431.15\%$ and $371.43\%$ ECE improvement compared to RR in medium-scale configuration under medium variance scenario and large-scale configuration under high variance scenario, respectively.
H$_2$O-Cloud and HDRL also yield up to $70.13\%$ and $75.39\%$ EE improvement compared to RR in small-scale configuration under high variance scenario and small-scale configuration under low variance scenario, respectively.
In regards to soft-deadline violation corresponding to QoS, H$_2$O-Cloud outperforms RR by up to $76.47\%$ in medium-scale configuration under medium variance scenario, while achieving lowest deadline violation rate $5.36\%$ in all experiments presented in this work.
On average, H$_2$O-Cloud outperforms RR by $329.35\%$, $39.77\%$, and $57.32\%$ in terms of Norm. ECE, Norm EE, and ddlVR, respectively.
\subsection{H$_2$O-Cloud and HDRL vs. DRL-Cloud} \label{exp:HHvsDRL}
DRL-Cloud \cite{cheng2018drl} comprehensively solves the energy cost reduction problem for large-scale CSPs by using a two-stage RP-TS processor. 
The first stage of RP-TS allocates a task to server farm and determines the timeslot to start processing the task, then the second stage chooses the exact server to run the task.

The difference between H$_2$O-Cloud and DRL-Cloud can be expressed in five aspects:
\subsubsection{Fine-grained Task Scheduling} H$_2$O-Cloud realizes a fine-grained scheduling by determining the exact minute(s) to run task, whereas in case of DRL-cloud, the task scheduling level of granularity is hourly. Fine-grained scheduling leads to better energy cost efficiency and energy efficiency.
\subsubsection{Hybridity} Both H$_2$O-Cloud and DRL-Cloud rely on DRL decision-making core to schedule tasks, however, H$_2$O-Cloud is embedded with a back-up hybrid mechanism, which functions whenever DRL cannot make right decisions in some corner cases. 
\subsubsection{Task Recycling Mechanism} \label{lowrejection} H$_2$O-Cloud defines task deadlines as soft, which means tasks can be recycled and rescheduled when the first scheduled processing time failed to meet the deadline. DRL-Cloud defines task deadlines as hard instead, and rejects tasks immediately when the hard-deadline is violated. The soft-deadline, task recycling mechanism, and hybridity of H$_2$O-Cloud yield lower task rejection rate.
\subsubsection{QoS-Awareness} H$_2$O-Cloud retrieves more user information to construct the user workload model, one of the important new parameters is priority. Priority defines the task importance. If the scheduled processing time range matches the priority, a reward will be given to the scheduling system as a user feedback (which is defined in Equation \ref{layer4reward}). H$_2$O-Cloud then uses the feedback to tune the scheduling system further in Layer 4.
\subsubsection{Professionalized Hierarchical Structure} H$_2$O-Cloud utilizes a more professionalized structure to accomplish the decision-making process hierarchically, i.e., one DQN per decision stage, while DRL-Cloud uses two DQNs for whole RP-TS process. Training each DQN individually provides it with specialized decision-making ability, which allows each DQN to concentrate on a specific decision step, which yields better scheduling decisions.

\begin{table*}[!h]
	\centering
	\vskip -0.5cm
	\caption{Improvement of HDRL and H$_2$O-Cloud Compared to DRL-Cloud under Low, Medium and High Variance Scenarios in Small, Medium, and Large Scales}\label{tbl:vsdrl}
	\resizebox{1\textwidth}{!}{
		\begin{tabular}{c|c|ccc|ccc|ccc}
			\hline
			\multicolumn{2}{c}{} & \multicolumn{3}{|c|}{Low Variance} & \multicolumn{3}{c}{Medium Variance} & \multicolumn{3}{c}{High Variance}\\\hline
			Scale & Approach & Norm. ECE & Norm. EE & Reward Rate & Norm. ECE & Norm. EE & Reward Rate & Norm. ECE & Norm. EE & Reward Rate \\\hline
			\multirow{2}{*}{Large}
			& HDRL				   & 68.73\% & 0.91\% & 285.71\% 
			& 34.15\% & 1.37\% & 499.25\%
			& 24.71\% & -          &458.93\% \\
			&H$_2$O-Cloud   & 91.06\% & 1.20\% & 289.22\% 
			& 48.63\% & 5.79\% & {551.76}\%
			& 26.93\% & -          & 450.22\% \\\hline
			\multirow{2}{*}{Medium}
			& HDRL				   & 80.50\% & 3.62\% & 157.45\% 
			& 76.32\% & 2.28\% & {538.33}\%
			& 45.24\% & 1.43\% & 264.71\% \\
			&H$_2$O-Cloud   & 75.15\% & 3.07\% & 214.66\% 
			& \bf{201.17}\% & 47.88 \% & 351.50\%
			& 65.97\% & 9.54\% & 216.12\% \\\hline
			\multirow{2}{*}{Small}
			& HDRL				   & 66.48\% & 4.22\% & 157.49\% 
			& \bf{85.90}\% & 23.97\% & 182.92\%
			& 56.24\% & 8.84\% & 227.54\% \\
			&H$_2$O-Cloud   & 30.63\% & -          & 270.50\% 
			& 79.76\% & 19.01\% & 167.95\%
			& 110.50\% & 29.43\% &25.30\% \\\hline		
		\end{tabular}
	}
\end{table*}

\begin{figure*}[!h]
	\centering
	\includegraphics[width=1\textwidth]{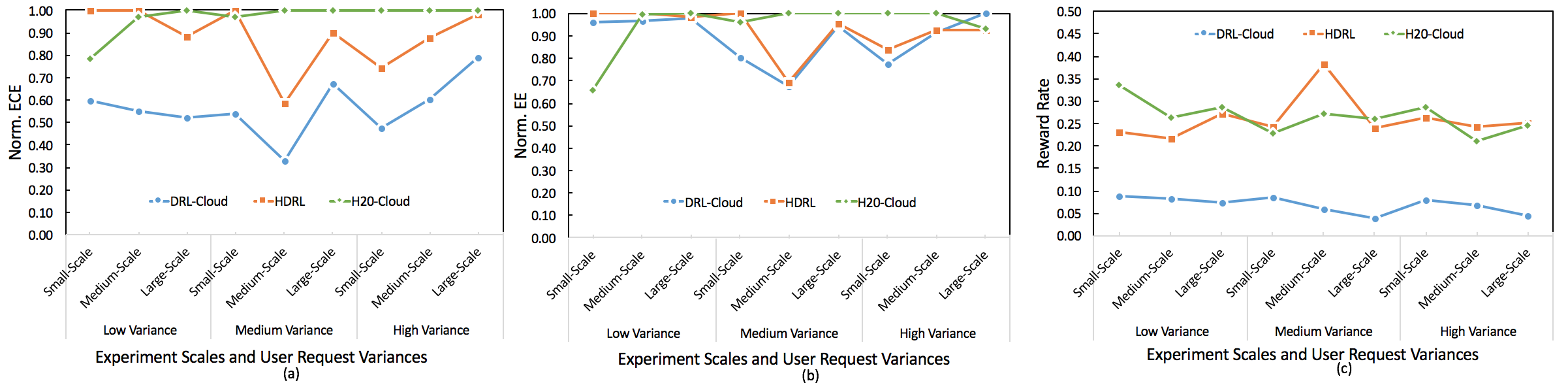}
	\vskip -4mm
	\caption{Comparison with DRL-Cloud under low, medium, and high variances scenario in small, medium, and large scales. Comparisons regarding Norm. ECE, Norm. EE, and Reward Rate are shown in (a), (b) and (c), respectively.}
	\label{fig:vsdrl}
	\vskip -5mm
\end{figure*}

The experiment results of comparison between H$_2$O-Cloud, HDRL and DRL-Cloud are shown in Figure \ref{fig:vsdrl} and Table \ref{tbl:vsdrl}. ECE and EE are normalized regarding highest values in each scenario.
Compared to DRL-Cloud, H$_2$O-Cloud and HDRL achieve higher ECE and Reward Rate in all scenarios as shown in Table \ref{tbl:vsdrl} and Figure \ref{fig:vsdrl} (a) and (c), respectively.
Higher scheduling precision (fine-grained scheduling) and professionalized hierarchical structure contribute the most to this result.
H$_2$O-Cloud and HDRL outperform DRL-Cloud regarding the Reward Rate because both of which are QoS-aware. Utilizing user feedback and introducing task priority help the scheduling system generate decisions that match the user preference as much as possible. 
As we can see in Figure \ref{fig:vsdrl} (b) and Table \ref{tbl:vsdrl}, H$_2$O-Cloud and HDRL outperforms DRL-Cloud regarding EE most of the time. This result shows that high ECE doesn't necessarily come from high EE. H$_2$O-Cloud and HDRL are trained to reach high ECE and high Reward Rate, so they may not outperform DRL-Cloud in terms of EE all the time.
On average, H$_2$O-Cloud outperforms DRL-Cloud with $81.08\%$ Norm. ECE and $281.91\%$ Reward Rate improvement, and HDRL outperforms DRL-Cloud with $59.81\%$ Norm. ECE and $308.04\%$ Reward Rate improvement.
The rejection rate of both H$_2$O-Cloud and HDRL ranges between $0\%$ and $0.1\%$ (the rejection rate of H$_2$O-Cloud is lower than HDRL), whereas the the rejection rate of DRL-Cloud ranges between $0.01\%$ to $1\%$.
Lower rejection rate of H$_2$O-Cloud is achieved by hybridity and task recycling mechanism as described in Section \ref{lowrejection}.

\subsection{H$_2$O-Cloud vs. HDRL} \label{exp:HvsH}
The rejection rates of both HDRL and H$_2$O-Cloud range between $0\%$ and $0.1\%$ in our various setup cases. It is noted that HDRL's goal in this work is to maximize energy cost efficiency, hence it learns to shut down idle servers so that working servers achieve good utilization rate. This means HDRL's rejection rate is higher than that of H$_2$O-Cloud (details appear in Section \ref{state_action_reward}). 

H$_2$O-Cloud outperforms HDRL in large-scale configuration and high variance scenario in terms of resource efficiency metrics, such as normalized ECE and normalized EE, and ddlVR. The reasons are twofold: 
i) For small-scale and medium-scale configurations, action space is much smaller than that of the large-scale configuration. 
This enables DRL agent to make intelligent decisions by fully learning the historical decisions and current status of environment. 
Note that in this case, the backup mechanism of RR in H$_2$O-Cloud would not help with resource efficiency. 
HDRL outperforms H$_2$O-Cloud in small-scale in terms of EE and ECE in Figure \ref{fig:vsdrl} because of the uncertainties in training.
ii) H$_2$O-Cloud consistently outperforms baselines in high variance scenario because the training of DRL agent is on-the-go and may introduce delay between changing environment and DQN. 
While environment is changing too fast for the DRL agent to keep up, decision of DRL agent will be less valid than RR agent, and that is where the hybrid nature of H$_2$O-Cloud improves the results (as shown by comparing the results with those of HDRL).

The comparison of H$_2$O-Cloud and HDRL in terms of Reward Rate is shown in Table \ref{tbl:vsdrl} and Figure \ref{fig:vsdrl} (c): We note that H$_2$O-Cloud does not consistently outperform HDRL, in terms of reward rate.
This is because the Reward Rate is not affected by the hybrid mechanism of H$_2$O-Cloud, which is the only difference between H$_2$O-Cloud and HDRL.
The up-and-downs shown in Figure \ref{fig:vsdrl} (c) are the result of the uncertainties in training process in both HDRL and H$_2$O-Cloud.
However, as noted before, H$_2$O-Cloud outperforms HDRL in terms of other design metrics, such as EE, ECE, and ddlVR.

\section{Conclusion}
In this work, H$_2$O-Cloud, a hierarchical and hybrid online cloud resource management and task scheduling framework, is presented to improve resource efficiency for warehouse-scale cloud service provider, while maintaining quality of service.
By utilizing hierarchy, and hybridity in system modeling and optimization, H$_2$O-Cloud achieves high performance in resource utilization, while maintaining robustness of framework.
H$_2$O-Cloud improves resource efficiency and quality of service remarkably, when compared to the baseline approaches, namely Round-Robin and DRL-Cloud. 
The experiment results confirm that the hierarchical structure with hybrid mechanism is effective and efficient in resource management and task scheduling for warehouse-scale cloud service providers.
H$_2$O-Cloud outperforms baseline HDRL (Hierarchical DRL) in large-scale configuration and high variance scenario in terms of energy cost efficiency, energy efficiency, and soft-deadline violation rate. This result further shows the improvements contributed by hybridity on top of the hierarchical DRL. 

\section*{Acknowledgement}
The authors gratefully acknowledge the support by the U.S. Army Research Office (ARO) under Grant No. W911NF-17-1-0076, the Defense Advanced Research Projects Agency (DARPA) Young Faculty Award under Grant No. N66001-17-1-4044 support, the National Science Foundation Career award under Grant No. CPS/CNS-1453860, and the National Science Foundation (NSF) Grant CCF-1423624. The views, opinions, and/or findings contained in this article are those of the authors and should not be interpreted as representing the official views or policies, either expressed or implied by the Defense Advanced Research Projects Agency, the Department of Defense or the National Science Foundation.

\ifCLASSOPTIONcaptionsoff
  \newpage
\fi



%



\bibliographystyle{IEEEtran}
\bibliography{bare_jrnl}
\vskip -14mm
\begin{IEEEbiography}[{\includegraphics[width=1in,height=1.2in,clip,keepaspectratio]{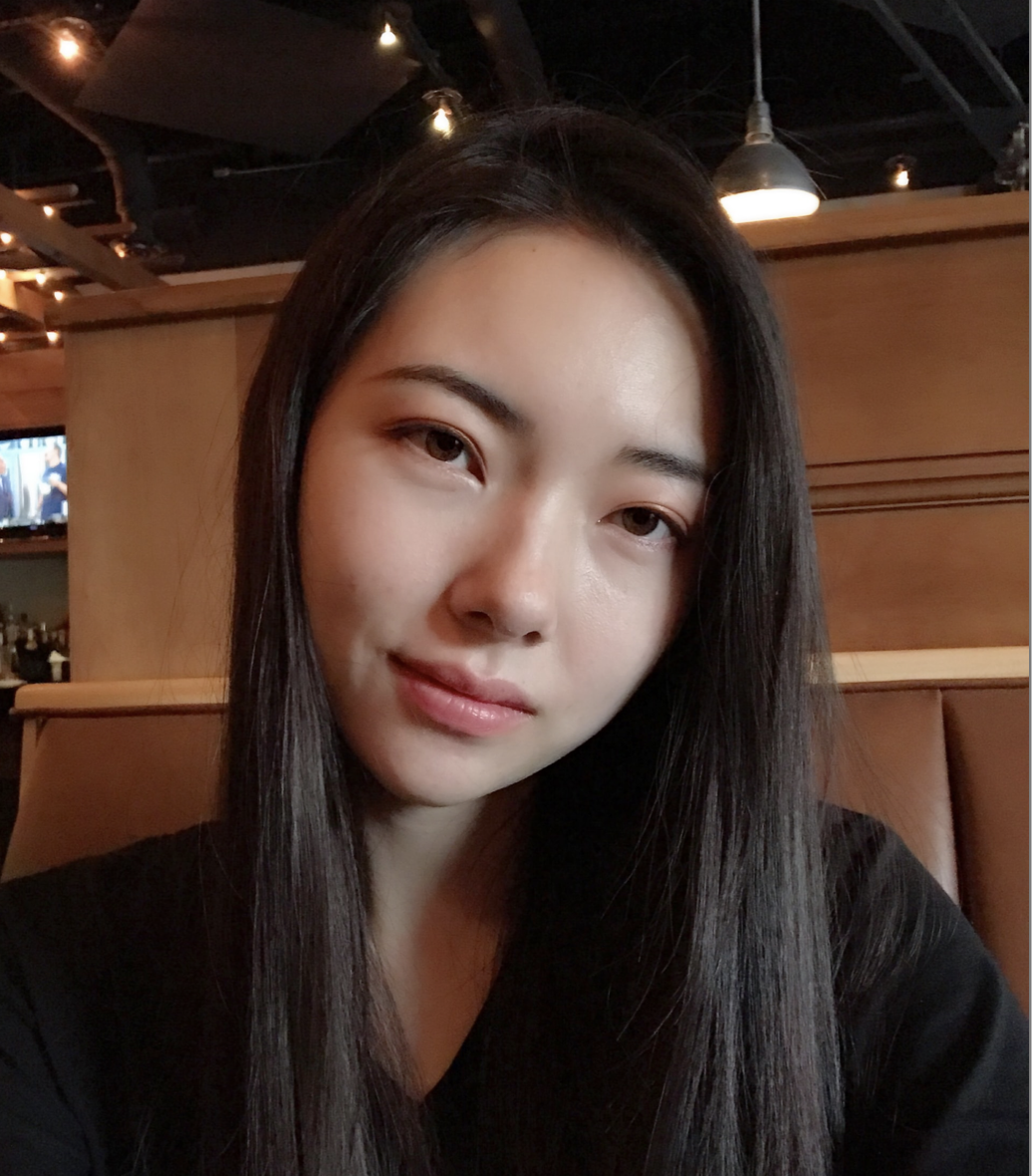}}]{Mingxi Cheng}
received her B.S. degree from Beijing University of Posts and Telecommunications (BUPT), China in 2016 and M.S. degree from Duke University in 2018. She is currently a Ph.D. student at University of Southern California (USC), under supervision of Prof. Nazarian and Prof. Bogdan. Her research interests include deep learning, deep reinforcement learning, and artificial intelligence.
\end{IEEEbiography}
\vskip -12mm
\begin{IEEEbiography}[{\includegraphics[width=1in,height=1.25in,clip,keepaspectratio]{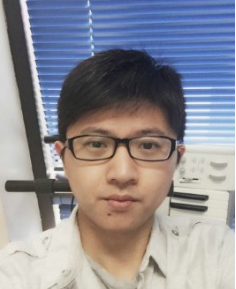}}]{Ji Li} received the B.S. degree in microelectronics from Xi’an Jiaotong University, Xi’an, China, in 2012, and the first M.S. degree, in 2014, in electrical engineering, the second M.S. degree, in 2017, in computer science, and the Ph.D. degree in electrical engineering under the supervision of Prof. S. Nazarian and Prof. J. Draper from the University of Southern California, Los Angeles, CA, USA. 
His research interests include deep learning, natural language processing, speech recognition. He is currently with Microsoft, Sunnyvale, CA, USA, researching on deep learning and natural language processing for intelligent service on Azure cloud.
\end{IEEEbiography}
\vskip -11mm
\begin{IEEEbiography}[{\includegraphics[width=1in,height=1.5in,clip,keepaspectratio]{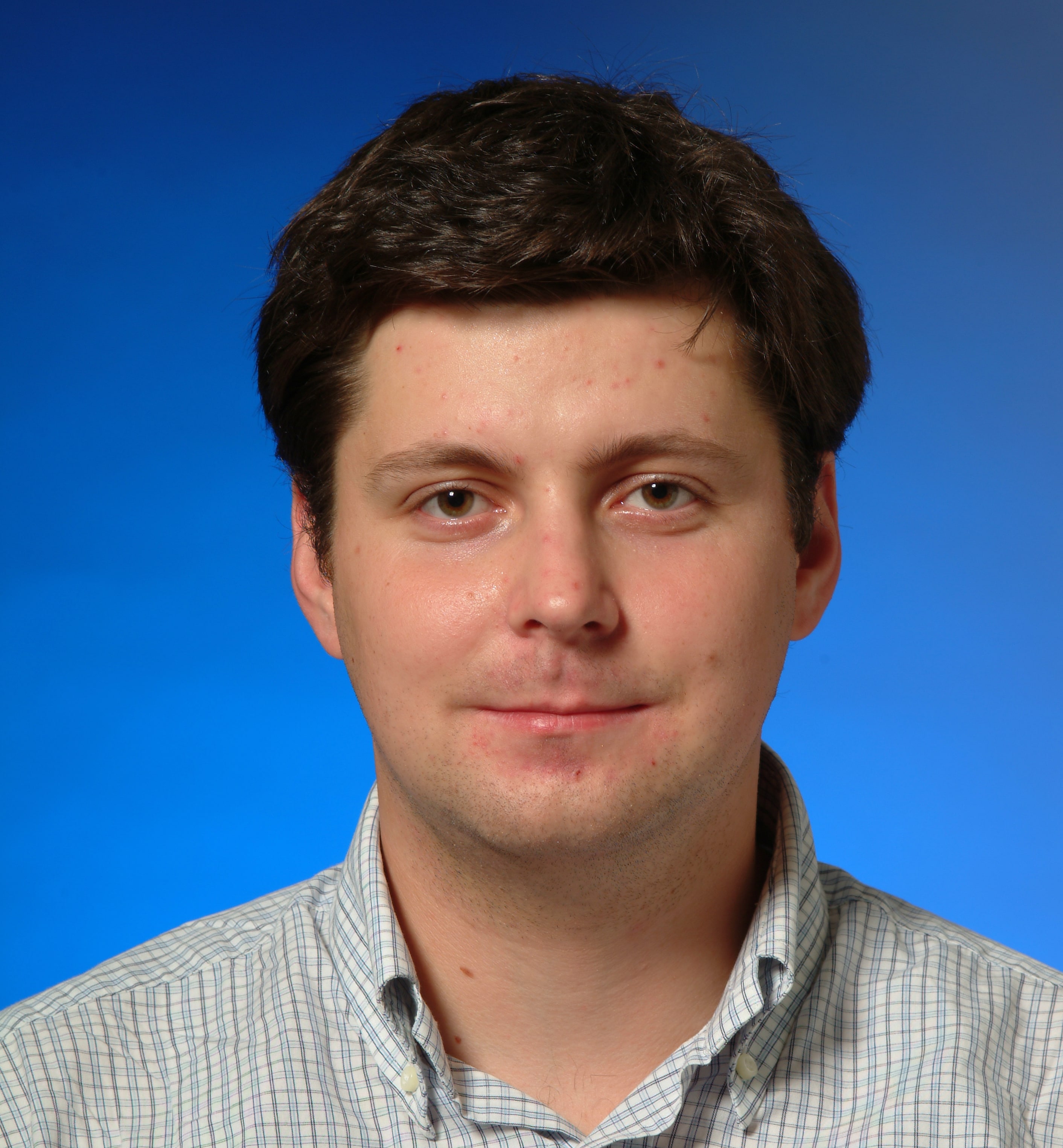}}]{Paul Bogdan} is an Associate Professor in the Ming Hsieh Department of Electrical and Computer Engineering at University of Southern California. He received his Ph.D. degree in Electrical \& Computer Engineering at Carnegie Mellon University. His work has been recognized with a number of honors and distinctions, including the IEEE CEDA Ernest S. Kuh Early Career Award, Defense Advanced Research Projects Agency (DARPA) Young Faculty Award, Okawa Foundation Award, National Science Foundation (NSF) CAREER Award, the 2013 Best Paper Award from the 18th Asia and South Pacific Design Automation Conference, the 2012 A.G. Jordan Award from Carnegie Mellon University for an outstanding Ph.D. thesis and service, the 2012 Best Paper Award from the Networks-on-Chip Symposium (NOCS), the 2012 D.O. Pederson Best Paper Award from IEEE Transactions on Computer-Aided Design of Integrated Circuits and Systems, the 2012 Best Paper Award from the International Conference on Hardware/Software Codesign and System Synthesis (CODES+ISSS), and the 2009 Roberto Rocca Ph.D. Fellowship. His research interests include the theoretical foundations of cyber-physical systems, the control of complex time-varying interdependent networks, the modeling and analysis of biological systems and swarms, new control algorithms for dynamical systems exhibiting multi-fractal characteristics, modeling biological / molecular communication, the development of fractal mean field games to model and analyze biological, social and technological system-of-systems, performance analysis and design methodologies for manycore systems.
\end{IEEEbiography}
\vskip -12mm
\begin{IEEEbiography}[{\includegraphics[width=1in,height=1.5in,clip,keepaspectratio]{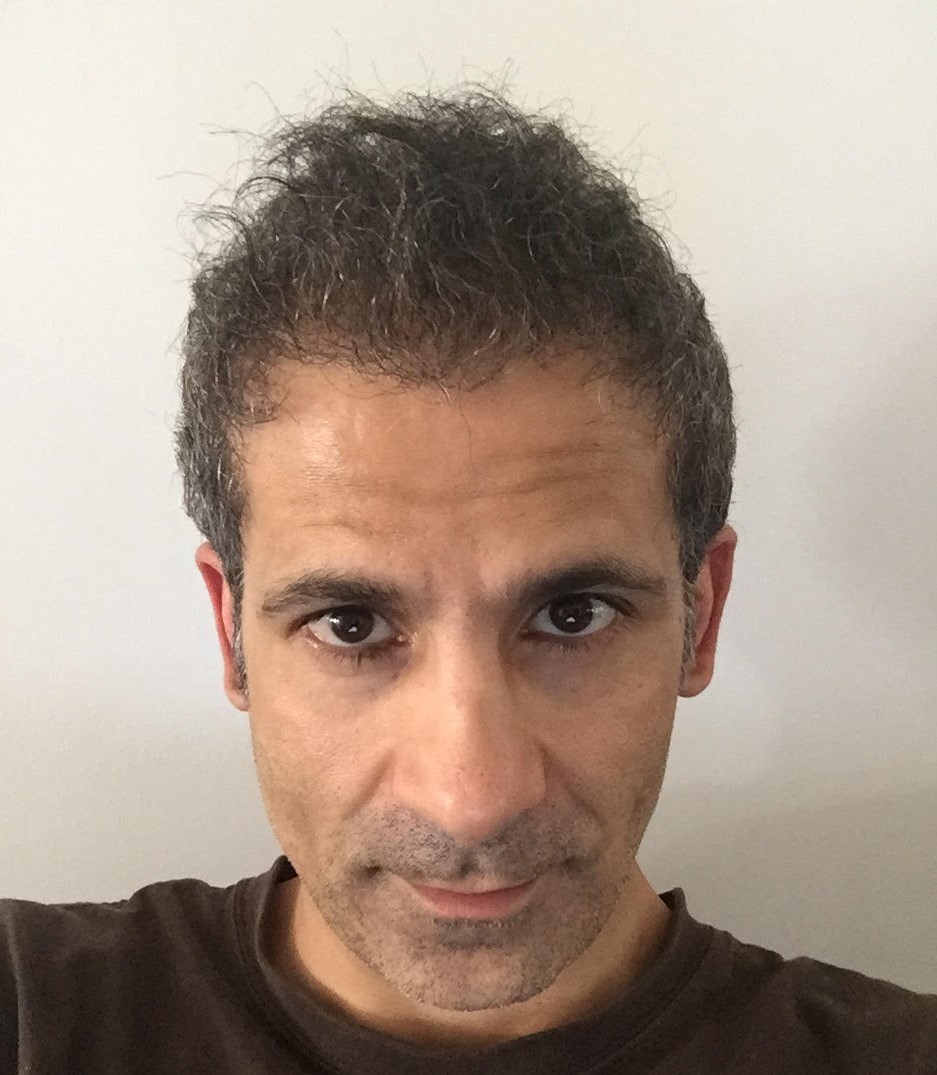}}]{Shahin Nazarian} is an Associate Professor of Engineering Practice in the Computer Engineering Division of the USC Viterbi School of Engineering. Prior to joining USC, he was a senior R\&D software engineer in Magma Design Automation (now part of Synopsys) focusing on timing and noise solution development as part of Talus and Tekton tools. He received his Ph.D. degree in Electrical Engineering from USC. His industrial experiences, as a consultant, technical expert, software and hardware design engineer, span over a wide range of areas including computer software analysis, computer architecture, and embedded systems. He is the first recipient of the Dean’s Award For Teaching Excellence. He is currently part of the Verification team of the ColdFlux project, developing CAD methodologies and tools for SFQ-based supercomputers. His research interests include computer-aided design, verification and optimization of system-on-chip, hardware acceleration, and signal integrity analysis.
\end{IEEEbiography}
\end{document}